\documentstyle[epsfig]{aipproc}
\begin{document}
\def \lsim            
{\raisebox{-3pt}{$\>\stackrel{<}{\scriptstyle\sim}\>$}}
\begin{flushright}
 UR-1626 \\
 ER/40685/963
\end{flushright}
\title{HERWIG for Top Physics\\ at the Linear Collider 
\footnote{Talk given 
by G. Corcella at Linear Collider Workshop 2000, Fermi National 
Accelerator Laboratory, Batavia, IL, U.\ S.\ A., 24-28 October 2000.}
}
\author{G. Corcella$^*$, E.K. Irish$^*$ and M.H. Seymour$^{\dagger}$}
\address{$^*$Department of Physics and Astronomy, University of Rochester,\\
Rochester, NY 14627, U.\ S.\ A.\\
$^{\dagger}$Rutherford Appleton Laboratory,\\
Chilton, Didcot, Oxfordshire. OX11 0QX, U.\ K.}
\maketitle
\begin{abstract}
We discuss recent improvement in the treatment of gluon radiation
in top production and decay in $e^+e^-$ processes according to
the HERWIG event generator and show studies on the top mass reconstruction
at the future Linear Collider.
\end{abstract}
For the sake of performing precision measurements of top quark properties 
at the future Linear Collider, trustworthy Monte Carlo simulations 
of multiparton radiation in top production and 
and decay will be essential. 
According to the standard algorithm of the HERWIG event 
generator \cite{herwig}, we shall refer to hereinafter, 
multiple radiation is treated in the soft or collinear approximation and 
no emission is permitted in the so-called `dead zones', which correspond 
to hard and large-angle parton radiation. 
The HERWIG algorithm can be improved by applying matrix-element 
corrections:
the dead zone is populated by the use of the exact first-order matrix element
(`hard correction') and the ${\cal O}(\alpha_S)$ result is used in the 
already-filled region any time an emission is the `hardest so far'
(`soft correction') \cite{mike}. 

One of the new features of HERWIG 6 \cite{her6} 
consists of the implementation of 
matrix-element corrections to top decays, 
which have been shown to have
a relevant effect on jet observables at the threshold for top pair production
\cite{corsey}. 
As pointed out in \cite{corsey1}, 
matrix-element corrections to top production in
$e^+e^-$ annihilation, implemented following \cite{mike1}, 
still needed improvement, since mass effects are not systematically 
included in the dead zone boundary and in the soft correction.

\begin{figure}
\centerline{\resizebox{0.49\textwidth}{!}{\includegraphics{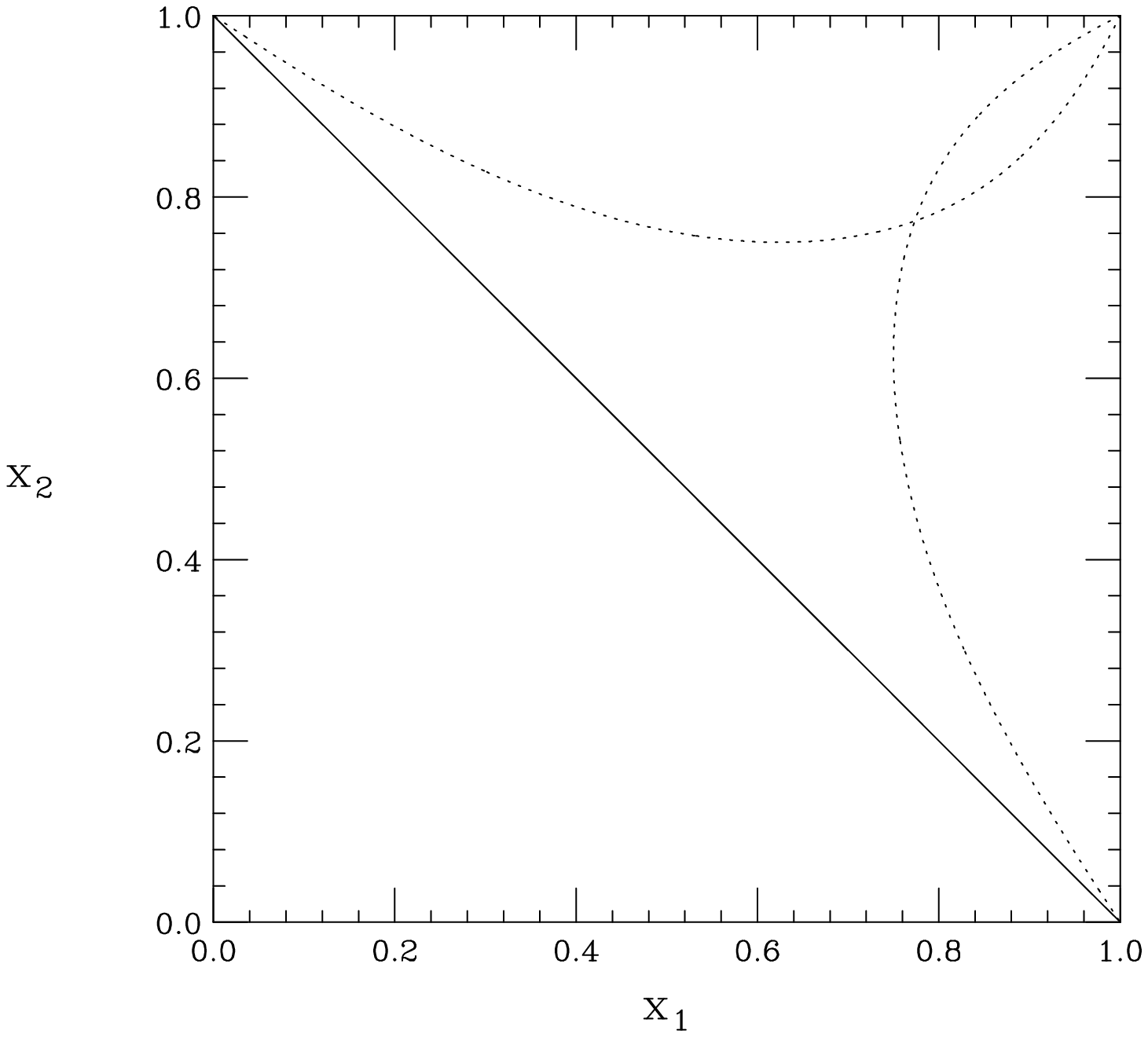}}%
\hfill%
\resizebox{0.49\textwidth}{!}{\includegraphics{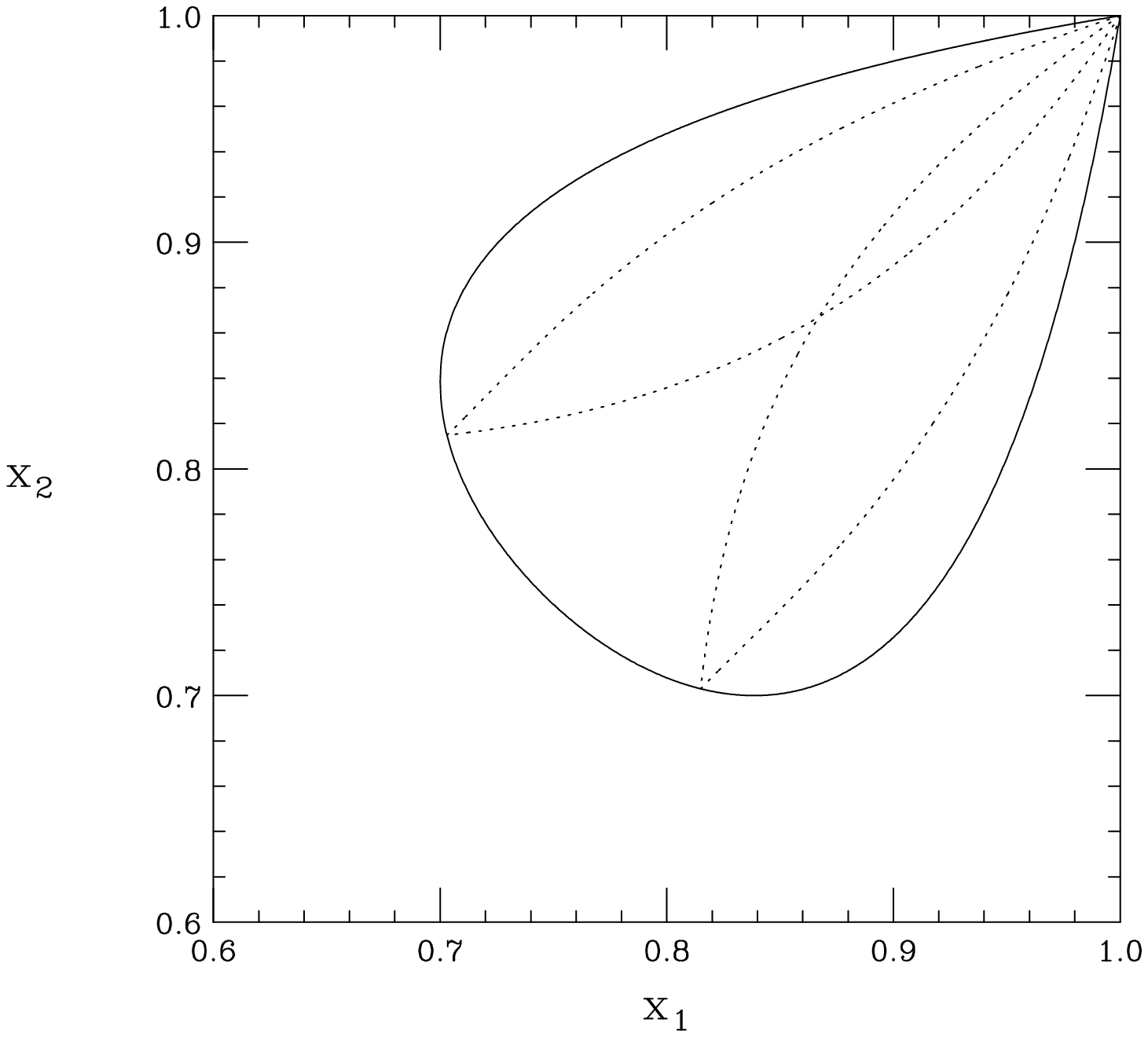}}}
  \caption{Total (solid) and HERWIG (dotted) phase-space limits for
massless quarks (left)
and for $t\bar t$ production (right) at $\sqrt s=500$~GeV.}
\label{fig1}
\end{figure}

In Fig.~\ref{fig1} we plot the total and HERWIG phase space 
for the process $e^+e^-\to q(p_1) \bar q(p_2) g(p_3)$, considering
massless quarks and top quarks at $\sqrt s =500$~GeV,
in terms of the energy fractions $x_1=2p_1\cdot q/q^2$ and
$x_2=2p_2\cdot q/q^2$, with $q=p_1+p_2+p_3$.   
We see that once we account for mass effects, the dead zone includes both a 
large- and a small-angle region of the physical phase space, 
the latter corresponding to 
the neighbourhood of the $x_1=x_2=1$ point, which, on the contrary, would be
entirely inside the HERWIG region if we neglected $m_t^2/s$ terms.
Since the soft singularity is not completely inside the HERWIG region, the
total emission into the dead zone, na\"\i vely calculated, 
would be infinite. As we did 
for the top-decay case \cite{corsey}, we avoid the soft singularity by setting 
a cutoff $E_{\mathrm{min}}$ on the energy of gluons which are 
radiated in the dead zone and check that phenomenological
observables are weakly dependent on the value of $E_{\mathrm{min}}$.
We choose $E_{\mathrm{min}}=2$~GeV as the cutoff default value.  

\begin{figure}
\centerline{\resizebox{0.49\textwidth}{!}{\includegraphics{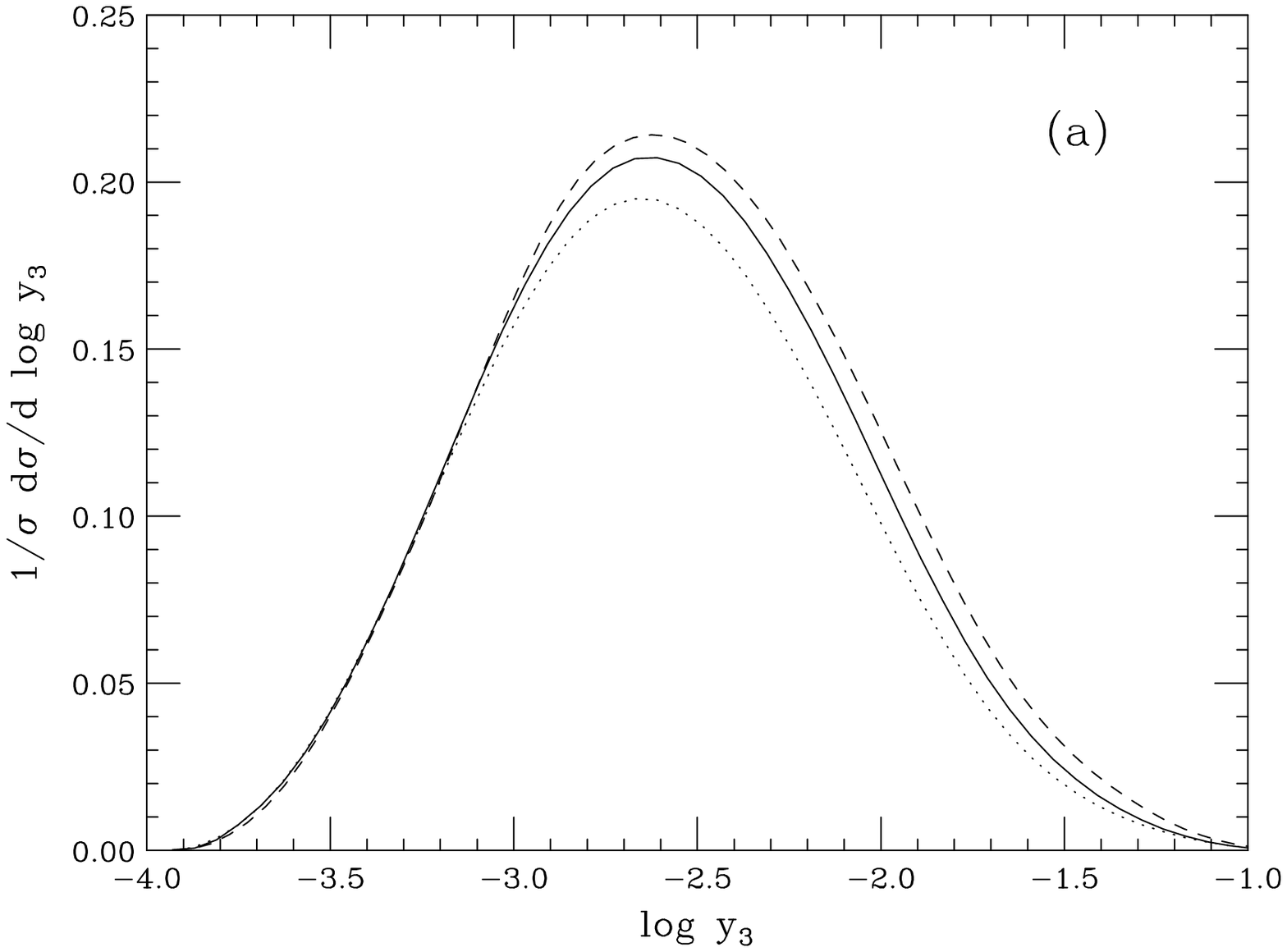}}%
\hfill%
\resizebox{0.49\textwidth}{!}{\includegraphics{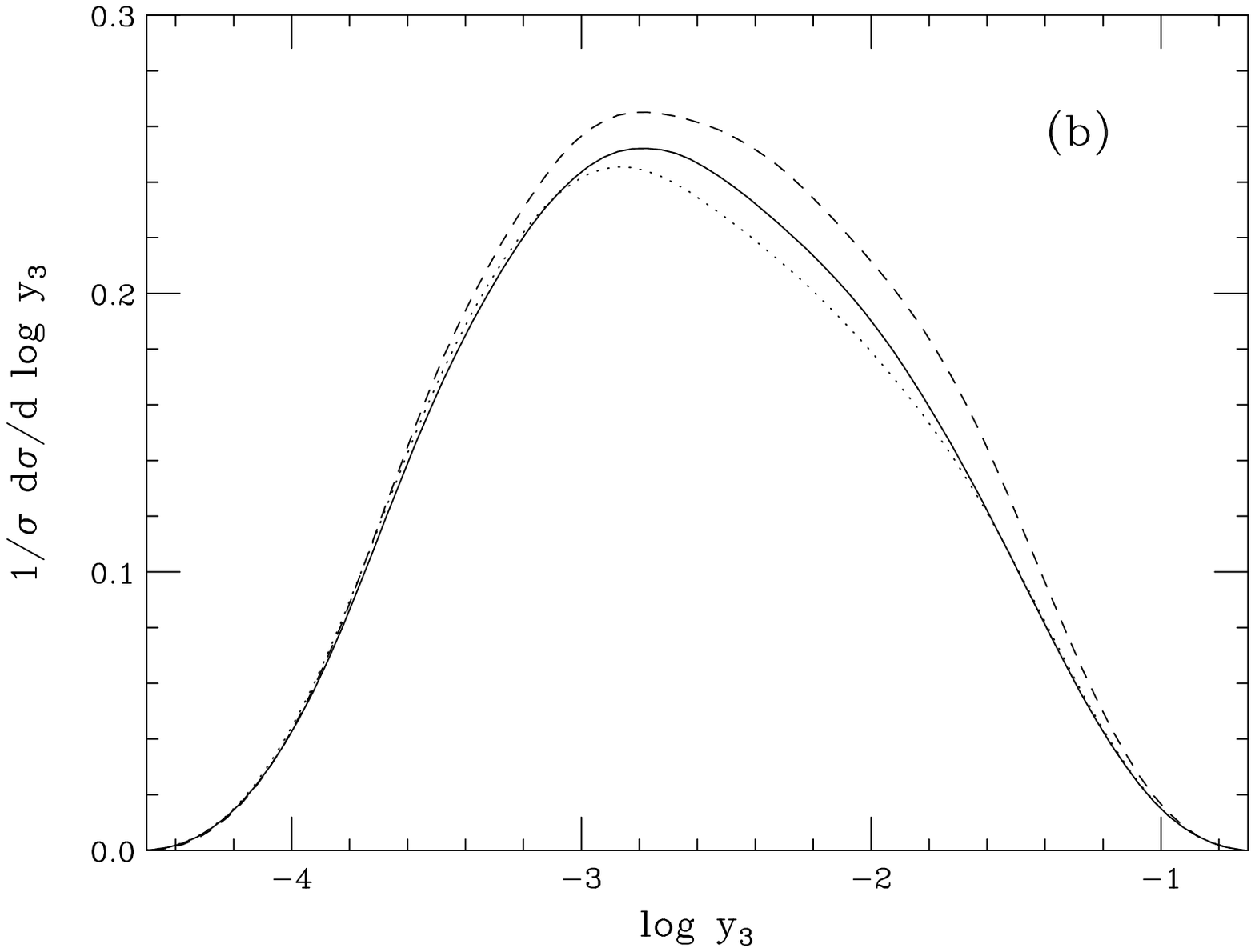}}}
  \caption{$y_3$ distributions at $\sqrt{s}=500$~GeV (a) and 
$\sqrt{s}=1000$~GeV (b)
according to HERWIG 6.2 (dashed line) and
6.3, once we populate either the small- and large-angle 
(solid) or only the large-angle region (dotted) of the dead zone.}
\label{fig2}
\end{figure}

We consider $e^+e^-\to t\bar t$ processes at $\sqrt{s}=500$~GeV and $1000$~GeV 
and cluster final-state partons into three jets by the use of the Durham 
algorithm \cite{catani}, assuming that both $W$'s decay leptonically.
We set the cuts $E_T>10$~GeV and $\Delta R>$~0.7 on transverse energy
and invariant opening angle of clustered jets. 
In Fig.~\ref{fig2} we plot the distributions of $y_3$, the threshold
value of the Durham variable for all events to be three-jet-like,
according to HERWIG 6.2, the latest public version, and
6.3, the new version in progress which will fully include 
mass effects in matrix-element corrections to top production.
We investigate the options to fill either the
small- and large-angle dead zone or only the large-angle region. 
The impact of the full implementation of $m_t^2/s$ effects 
is a suppression of emission, which is more visible 
at $\sqrt{s}=1000$~GeV, as the radiation in the top-production stage gets
more important. Filling the small-angle region
as well results in more events at intermediate values of $y_3$.
We checked that once the centre-of-mass
energy is increased so that terms $m_t^2/s$ are negligible, the
6.3 results reproduce the 6.2 ones. 

Matrix-element corrections to $W\to q\bar q'$ decays in the top decay, 
not yet included in \cite{her6}, turn out to be a 
straightforward extension of the corrections to $Z\to q\bar q$ processes
in the massless approximation $m_{q,q'}\ll m_{W,Z}$.
We found little impact on generic jet observables
at the Linear Collider, even at the top threshold,
where the radiation in the production phase is negligible.
In fact, the already-existing jets associated with
the $b$ quarks from the top decay and with the $W$-decay products,
even in the soft or collinear approximation, make the detection
of hard and large-angle gluon radiation in the $W$ decay pretty difficult.
Systematic and detailed analyses to find out possible variables 
which might be sensitive to matrix-element corrections to $W$ decays are 
in progress.

\begin{figure}
\centerline{\resizebox{0.49\textwidth}{!}{\includegraphics{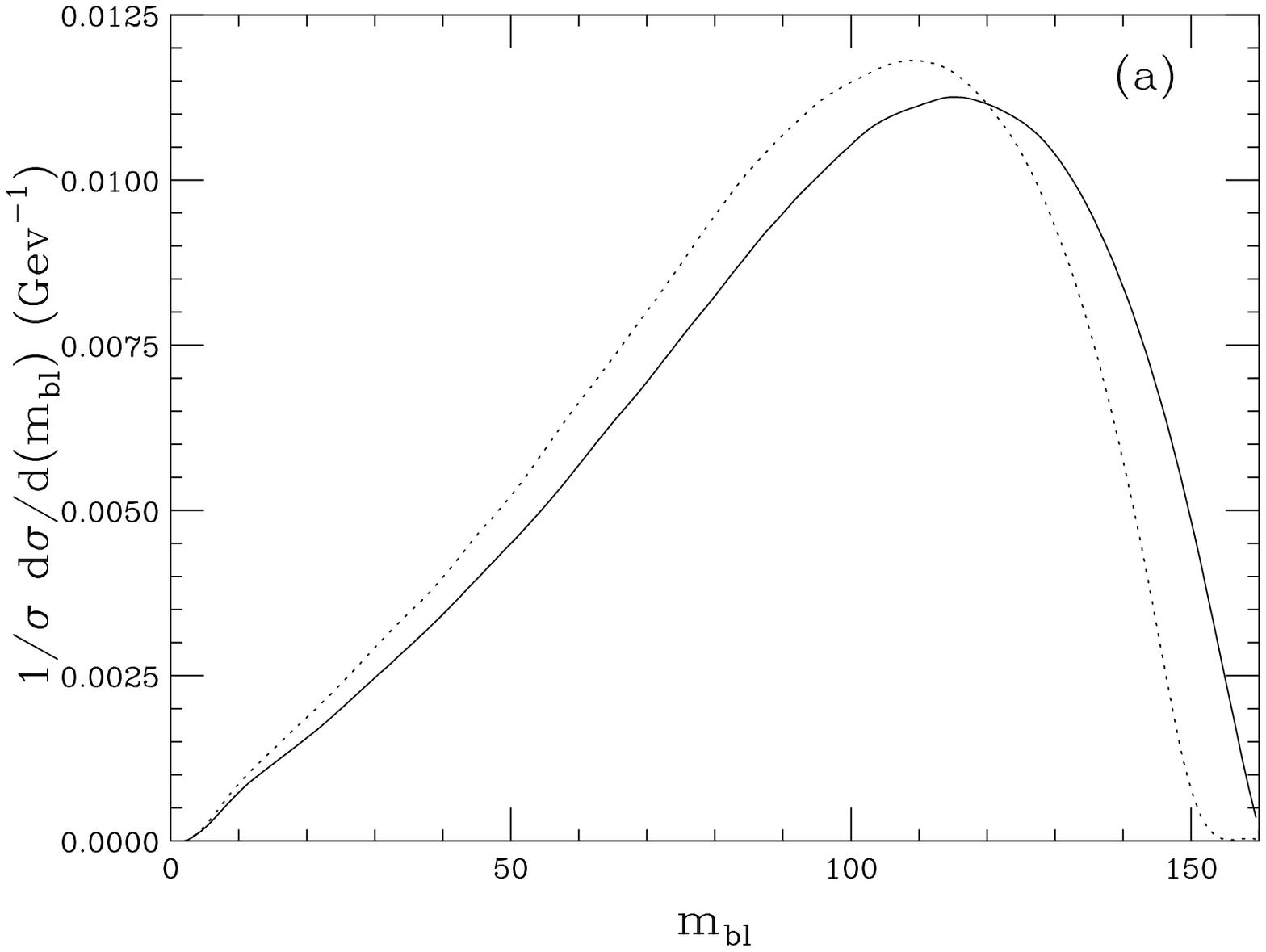}}%
\hfill%
\resizebox{0.49\textwidth}{!}{\includegraphics{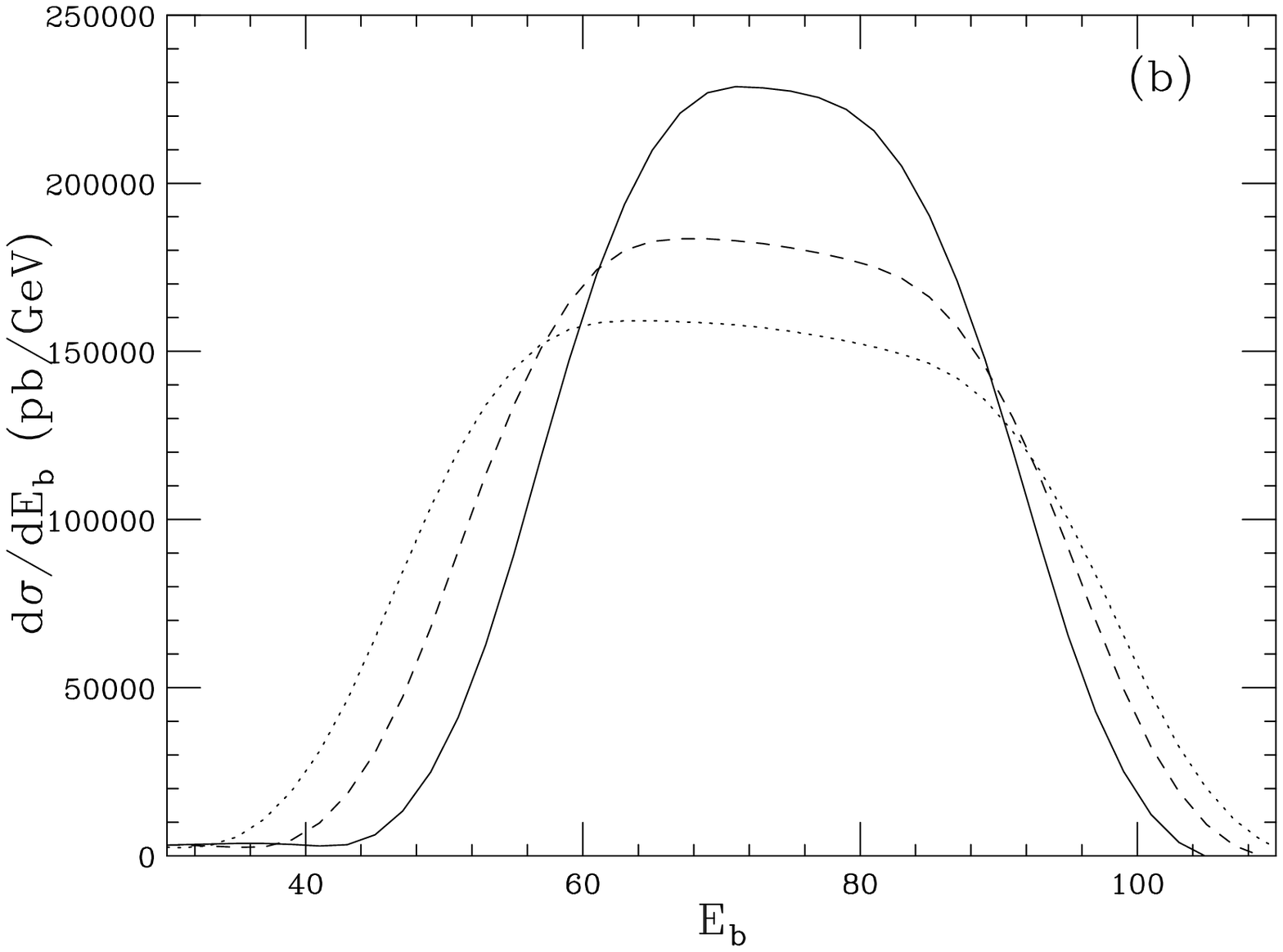}}}
  \caption{(a): Distribution of the invariant mass $m_{b\ell}$ 
for $m_t=171$~GeV (dotted line) and $m_t=179$~GeV (solid). 
(b): $b$ energy $\sqrt s =370$~GeV for $m_t=179$ (solid),
175 (dashed) and 171 (dotted) GeV.}
\label{fig4}
\end{figure}

We finally wish to report on studies on the top mass reconstruction in the 
dilepton channel. We consider 
the $b$-lepton invariant mass $m_{b\ell}$ and the $b$ energy
$E_b$, where the $b$ quark is considered together with its gluon radiation,
as possible variables which may allow a fit
of the top mass. Being a Lorentz-invariant observable, the 
$m_{b\ell}$ distribution is independent of the centre-of-mass energy and,
as already pointed out in \cite{cms} for the purpose of hadron collisions,   
of the hard-scattering process as well. In fact, within the 
statistical Monte Carlo fluctuations, we find the same results for different 
values of $\sqrt{s}$. On the contrary, $E_b$ is not 
Lorentz-invariant, hence it will be sensitive to the boost from the top rest
frame, where the top decay is performed, to the laboratory frame. 
At the threshold for $t\bar t$ 
production, the dependence of $E_b$ on the top mass will be emphasized,
as the $t\bar t$ pair is produced almost at rest.
Fig.~\ref{fig4} shows that the $m_{b\ell}$ distribution is shifted towards
larger values as the top mass is increased. Moreover, 
the half-maximum width $\sigma_b$ of the 
$E_b$ distribution shows a strong dependence on the top mass at 
$\sqrt{s}=370$~GeV, since the distribution gets narrower as the top mass
approaches the threshold value $\sqrt s/2$.
If we try to parametrize the relation of the average value 
$\langle m_{b\ell}\rangle$ and 
$\sigma_b$ in terms of $m_t$, we find that the best fits
are a straight line for $\langle m_{b\ell}\rangle$ and 
a parabola for $\sigma_b$:
\begin{eqnarray}
\langle m_{b\ell}\rangle &=&\ \ 0.756\ m_t-37.761\ {\mathrm GeV}\ ;\\
\sigma_b &=&-0.081\ m_t^2 +26.137\ m_t-2048.968\ {\mathrm GeV}\ .
\label{fits}
\end{eqnarray}
Inverting Eqns.~(1) and (2) to extract $m_t$, we conclude that if 
$\Delta \langle m_{b\ell}\rangle$ and 
$\Delta \sigma_b$ are the uncertainties on measurements of 
the invariant mass and of the half-maximum width, they will result in
an error $\Delta m_t\approx 1.32 \  \Delta \langle m_{b\ell}\rangle$ and
$\Delta m_t\approx 0.35-0.65\ \Delta \sigma_b$, where the latter 
uncertainty refers to the range 171 GeV $\lsim m_t\lsim$ 179 GeV.
The $b$ energy looks therefore to be a quite promising observable
with which to extract 
$m_t$, although we would need to know the experimental accuracy on the 
masurement of $\sigma_b$ in order to estimate the foreseen 
uncertainty on $m_t$. Furthermore, we expect that $E_b$ and
$\sigma_b$ will be sensitive to the beam energy smearing, which
has not been accounted for in the plots of Fig.~\ref{fig4}.
The implementation of beamsstrahlung in HERWIG, via an interface with 
the CIRCE program \cite{circe}, is under way.

In summary, we discussed recent progresses in the
implementation of  matrix-element corrections
to the HERWIG simulation of 
top production and decay at the Linear Collider and showed 
studies on the top mass reconstruction, for which purpose the 
$b$ energy is expected to be an interesting variable for 
$e^+e^-$ collisions slightly above the top threshold.

We acknowledge U. Baur, D.W. Gerdes and L.H. 
Orr for discussions on the top mass reconstruction.
The work of G.C. is 
supported by grant number DE-FG02-91ER40685 from the U.S. Dept. of Energy.
 
\end{document}